\documentclass[aps,prb,amsmath,amssymb,twocolumn,superscriptaddress,showpacs,floatfix]{revtex4-1}

\usepackage{ifpdf}
 \newif\ifpdf
\ifx\pdfoutput\undefined
   \pdffalse
\else
   \pdfoutput=1
   \pdftrue
\fi
\ifpdf
   \usepackage{graphicx}
   \usepackage{epstopdf}
\else
   \usepackage{graphicx}
\fi

\DeclareMathOperator{\TCFE}{\mathit{T}_{\mathrm{\scriptscriptstyle C}}^{\mathrm{\scriptscriptstyle FE}}}
\DeclareMathOperator{\TCFM}{\mathit T_{\mathrm{\scriptscriptstyle C}}^{\mathrm{\scriptscriptstyle FM}}}

\DeclareMathOperator{\TM }{\mathit T_{\mathrm{\scriptscriptstyle M}}}
\DeclareMathOperator{\TB }{\mathit T_{\mathrm{\scriptscriptstyle B}}}
\DeclareMathOperator{\Jtot }{\mathit J_{\mathrm{tot}}}
\DeclareMathOperator{\JJ }{\mathit J}
\usepackage{graphicx}% Include figure files
\usepackage{dcolumn}% Align table columns on decimal point
\usepackage{bm}% bold math

\begin{document}

%\preprint{}

\title{Coupling of ferroelectricity and magnetism through Coulomb blockade in Composite Multiferroics }

\author{O.~G.~Udalov}
\affiliation{Department of Physics and Astronomy, California State University Northridge, Northridge, CA 91330, USA}
\affiliation{Institute for Physics of Microstructures, Russian Academy of Science, Nizhny Novgorod, 603950, Russia}

\author{N.~M.~Chtchelkatchev}
\affiliation{Department of Physics and Astronomy, California State University Northridge, Northridge, CA 91330, USA}
\affiliation{L.D. Landau Institute for Theoretical Physics, Russian Academy of Sciences,117940 Moscow, Russia}
\affiliation{Department of Theoretical Physics, Moscow Institute of Physics and Technology, 141700 Moscow, Russia}

\author{I.~S.~Beloborodov}
\affiliation{Department of Physics and Astronomy, California State University Northridge, Northridge, CA 91330, USA}

\date{\today}

\begin{abstract}
Composite multiferroics are materials exhibiting the interplay of ferroelectricity,
magnetism, and strong electron correlations. Typical example ---
magnetic nano grains embedded in a ferroelectric matrix.
Coupling of ferroelectric and ferromagnetic degrees of freedom in these materials is due to the influence of
ferroelectric matrix on the exchange coupling constant via screening of the intragrain
and intergrain Coulomb interaction. Cooling typical magnetic materials the ordered state appears at
lower temperatures than the disordered state. We show that in
composite multiferroics the ordered magnetic phase may appear at higher temperatures than the
magnetically disordered phase. In non-magnetic materials such a behavior is known as inverse phase transition.

\end{abstract}


\maketitle

\section{Introduction.}

Currently composite materials with combined magnetic and electric degrees of freedom attract much of attention for their promise to produce new effects and functionalities~\cite{Scott2006,Spal2007,Bar2008}. The idea of using ferromagnetic and ferroelectric properties
in a single phase multiferroics was developing since seventeenths~\cite{Schol1974}. However,
in bulk homogeneous materials this coupling is
weak due to relativistic parameter $v/c$, with $v$ and $c$ being the electron velocity and the speed of light, respectively.
Only recently the new classes of two-phase multiferroic materials such as single domain multiferroic nanoparticles~\cite{Math2008}, laminates~\cite{Lin2004,Hayes2002}, epitaxial multilayers~\cite{Garg2013,Dkhil2010}, and granular materials~\cite{He2007,Park2006,Ram2004} were discovered giving a new lease of life to this field.
\begin{figure}
\includegraphics[width=1\columnwidth]{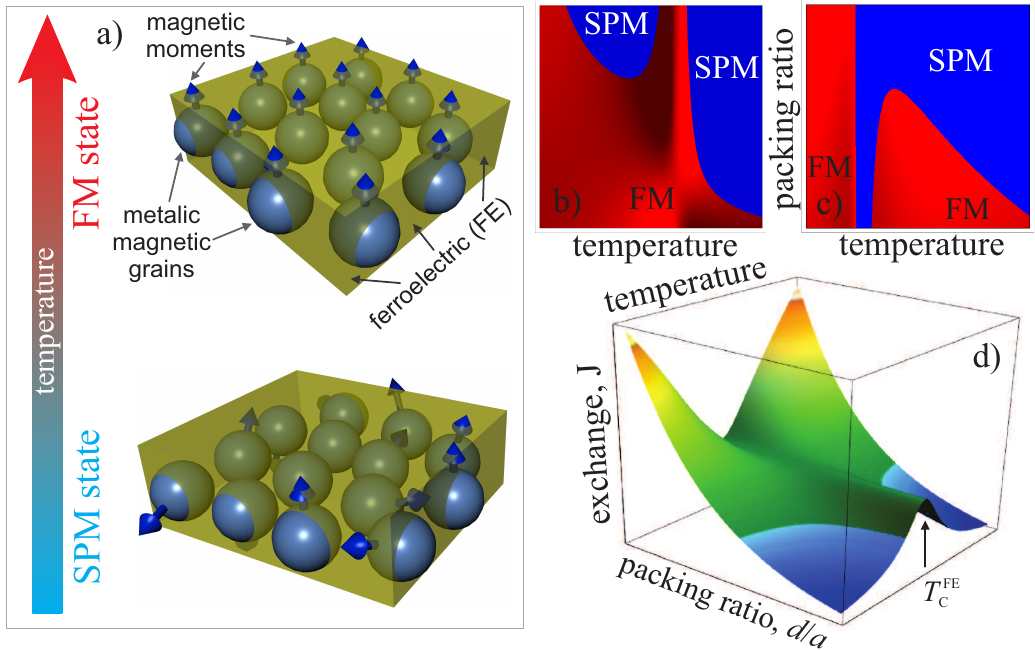}\\
\caption{(Color online) a) Sketch of a granular multiferroic (GMF) consisting of ferromagnetic grains with magnetic moments
embedded in a ferroelectric matrix. Cooling typical magnetic materials the ordered phase appears at lower temperature than the disordered phase. In composite multiferroics the ordered (FM) state may appear at higher temperature (upper panel) than
the disordered (SPM) state (lower panel).
In non-magnetic materials such a behavior is known as inverse phase transition. Diagrams b) and c)
show magnetic state of GMF in coordinates temperature, T vs. packing ratio, $d/a$, with $d$ and $a$
being the average intergrain distance and the grain size, respectively. The diagrams b) and c)
correspond to the limits of large and small intergrain distance $d$, respectively.
Graph d) shows the intergrain exchange constant $J$ vs. temperature $T$ and the packing ratio, $d/a$.}\label{fig1}
\end{figure}
So far, the interface strain generated by the ferroelectric layer was considered as the promising mechanism for strong enough magnetoelectric coupling in two-phase multiferroic materials~\cite{Band2010,Schroder1982,Scott2006,Ram2004,He2007}. This strain modifies the magnetization in the magnetic layer and the magnetic anisotropy energy.

We propose a different mechanism for magnetoelectric coupling emerging at the edge of strong long-range electron interaction, ferroelectricity, and magnetism. In composite multiferroics --- materials consisting of metallic ferromagnetic grains embedded into ferroelectric (FE) matrix, Fig.~\ref{fig1}, the origin of this coupling is twofold: i) Strong influence of FE matrix on the Coulomb gap defining the electron localization length and the overlap of electron wave functions, and therefore controling the exchange forces. ii) Dependence of the long-range part of Coulomb interaction, and thus the exchange interaction, on the dielectric permittivity of the FE matrix.

Granular magnets consist of nanosized single domain ferromagnetic particles. Each particle has uniform magnetization and its own non zero magnetic moment. Direction of a single particle magnetization and collective behavior of the particle ensemble depend on particle magnetic anisotropy and the interparticle interaction. For weak interparticle interaction and small anisotropy 
the magnetic moment of a single particle is not fixed and fluctuates in time. Magnetic moments of different particles are not correlated. This is so-called superparamagnetic (SPM) state.~\cite{Livin1959} Interparticle interactions (such as dipole-dipole,~\cite{Vinai1999,Trohidou1998} and exchange,~\cite{Hembree1996,Chen2010}) can lead to establishing of magnetic order with correlated magnetic moments of different particles. Due to interactions the disordered SPM system can come to the 
ordered ferromagnetic (FM) or antiferromagnetic state with decreasing temperature. We discuss in this paper the influence of ferroelectric matrix on the interparticle exchange interaction.

We show that the effective ferromagnetic exchange constant $J$ between the ferromagnetic grains strongly depends on temperature near the ferroelectric Curie temperature $\TCFE$ in granular multiferroics due to the above mentioned mechanisms. The transition temperature between ordered and disordered magnetic states can be found approximately using the equation $J(T) = T$. FM state corresponds to $J(T) > T$. If mechanism (i) is the strongest, the FM state appears at higher temperatures than the disordered SPM state, Fig.~\ref{fig1}. Such a behavior originates from the fast growth of the exchange coupling with temperature, $\frac{dJ}{dT}\gg 1$, in the vicinity of paraelectric-ferroelectric phase transition. This is known as an inverse phase transition. It appears in various systems such as He$^3$ and He$^4$, metallic alloys, Rochelle salt  ferroelectrics, polymers, and high-$T_c$ superconductors~\cite{Dobbs2002,Loidl2000,Kelle1999,Truskett2001,Axelos1997}. Here we predict the inverse phase transition in magnetic materials and address the main question of why the interplay of Coulomb blockade, ferroelectricity, and ferromagnetism
in granular multiferroics (GMF) leads to such a peculiar temperature dependence of the exchange coupling $J(T)$.

\section{Quantum nature of composite multiferroics.} 

Composite multiferroics are characterized by two temperatures: i) the ferroelectric Curie temperature $\TCFE$ describing the paraelectric-ferroelectric transition of FE matrix, and ii) the ferromagnetic grains Curie temperature, $\TCFM$. For temperatures $T > \TCFM$ the grains are in the paramagnetic state with zero magnetic moments.
For temperatures $T < \TCFM$ each grain is in the FM state with finite
spontaneous magnetic moment. The temperature $\TCFM$ depends on grain sizes. Here we assume that all grains have the same ordering Curie temperature $\TCFM$ with $\TCFM\gg \TCFE$.

Although each grain is in the ferromagnetic state for temperatures  $T < \TCFM$ the 
whole system has several types of magnetic behavior depending on the ratio of temperature and 
several energy scales: 1) the grain anisotropy energy $E_a$~\cite{Livin1959}, 2) the intergrain exchange coupling $J$~\cite{Hembree1996,Chen2010}, and 3) the magneto-dipole interaction $E_{d}$~\cite{Vinai1999,Trohidou1998}. For temperatures $T > {\rm max}(E_a, J, E_{d})$ the grain magnetic moments are uncorrelated and fluctuate in time. In this case the whole system is in the SPM state~\cite{Boz1972}. For temperatures $T$ below than one of the above energy scales the system magnetic state changes. Depending on the ratio of $E_a$, $J$, and $E_{d}$ the different states are possible~\cite{Chen2010}.

The grain anisotropy energy $E_a$ has two contributions coming from the grain bulk and grain surface.
The anisotropy axis varies from grain to grain due to the grain shape and disorder orientation.
For large anisotropy energy, $E_a > {\rm max}(J, E_d) $ and temperatures $T < E_a$ the grain moments are frozen and not correlated. The temperature $\TB = E_a$ is called the blocking temperature.

In this paper we consider the opposite case of large exchange energy, $J > {\rm max} (E_{a},E_{d})$, with negligible
bulk and surface magnetic anisotropies, $E_a$ and magneto-dipole interaction, $E_d$. This limit is realized for small grains~\cite{Freitas2007,Boz1972,Hel1981}. The magnetic phase transition occurs  
at temperatures $\TM = J$ in this case. The system moves from the SPM state with 
uncorrelated magnetic moments of grains to the FM state with co-directed spins of grains. The temperature $ \TM = J$ is called the ordering temperature. 
Below we study the influence of FE matrix on intergrain exchange interaction and on the ordering temperature $\TM$ of GMF.

Consider the exchange interaction of two metallic ferromagnetic grains of equal sizes, $a$. Each grain is characterized by the Coulomb energy $E_c=e^2/(a\epsilon)$ with $e$ and $\epsilon$ being the electron charge and the average dielectric permittivity of the granular system, respectively. We assume that the Coulomb energy is large, $E_c \gg T$ and the system is in the insulating state with negligible electron hopping between the grains. In this case the exchange interaction has a finite value due to the overlap of electron wave functions located in different grains.
\begin{figure}
  \includegraphics[width=0.93\columnwidth]{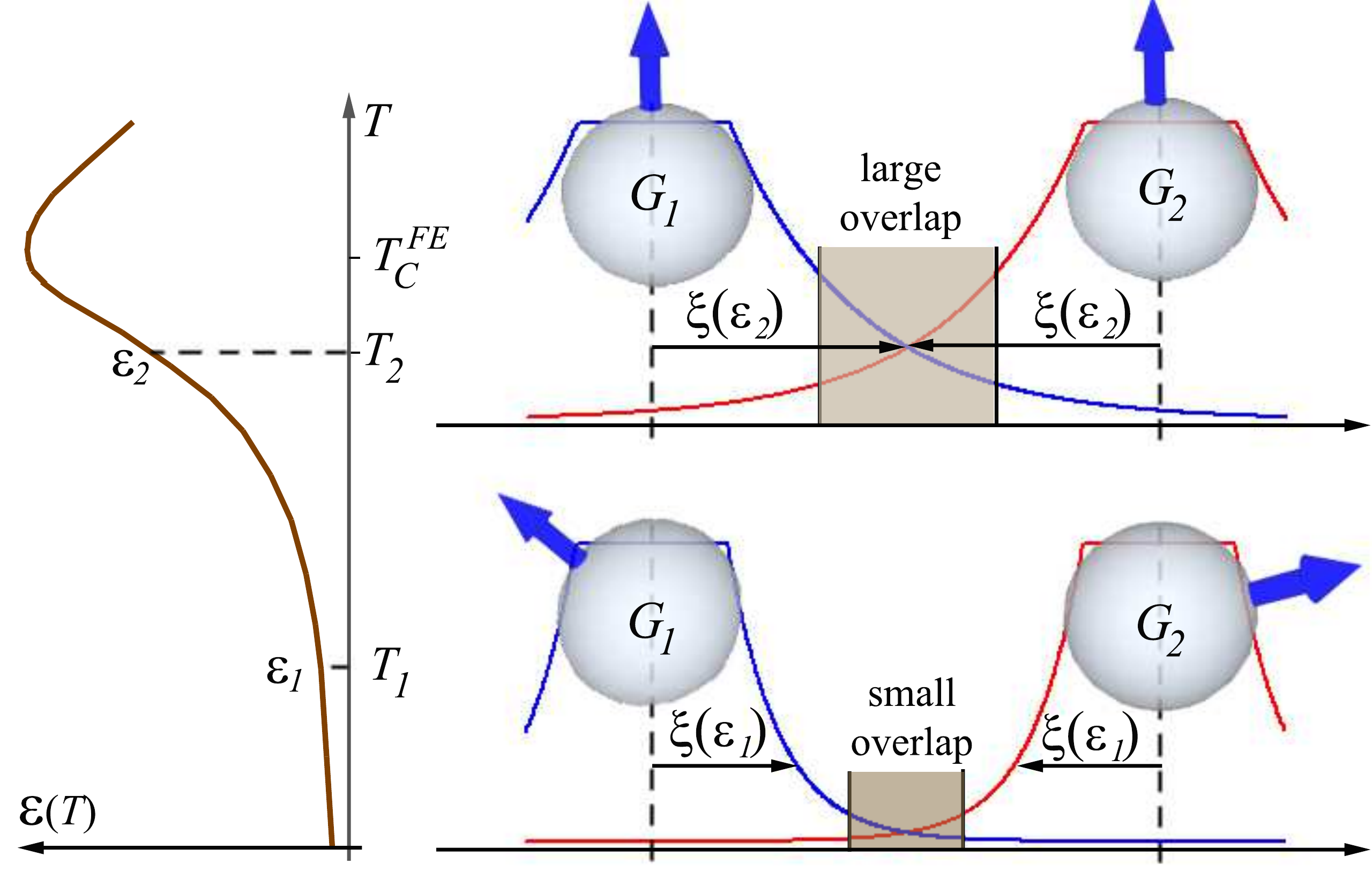}\\
  \caption{(Color online) Mechanism for magnetoelectric coupling in composite multiferroics.
  Right panel shows the overlap of the electron wave functions (blue and red curves) located
  in grains G$_1$ and G$_2$ embedded into ferroelectric (FE) matrix. This overlap defines the exchange
  coupling $J$ in Eq.~(\ref{Eq_exch}) and the strength of spin correlations (blue arrows).
  The localization length $\xi(\epsilon)$, with $\epsilon$ being the dielectric permittivity of FE matrix,
  shows the characteristic decay of electron wave functions.
  Left panel shows $\epsilon(T)$ vs. temperature $T$.
  The dielectric permittivity $\epsilon$ is small for temperatures $T = T_1 \ll T^{FE}_C$, with $T^{FE}_C$ being
  the FE transition temperature, leading to small localization length
  $\xi(\epsilon)$ and small overlap of electron wave functions resulting in a
  small exchange coupling $J$ and uncorrelated spin state.
  Close to the FE transition ($T=T_2$) the dielectric permittivity $\epsilon$
  is large leading to the large overlap of electron wave functions and to the
  strong exchange coupling resulting in the ferromagnetic state.}\label{Expl}
\end{figure}

We describe the coupling of each pair of electrons as $-J_{ij}(\hat{s}_i\cdot\hat{s}_j)$ with $\hbar\hat{s}$ being the spin operator with indexes $i$ and $j$ numbering electrons in the first and the second grain, respectively, and the parameter $J_{ij}$ being the exchange interaction of two electrons. The total exchange interaction of two grains can be written as a sum over all electrons, $\Jtot = -\sum_{ij}J_{ij}(\hat{s}_i\cdot\hat{s}_j)$. Below for simplicity we assume that $J_{ij} = J$ does not depend on indexes $i$ and $j$. Thus, the Hamiltonian has the form $\Jtot=-J(\hat{S}_1\cdot \hat{S}_2)$, where $\hbar\hat{S}_{1,2}$ is the total spin of the first (second) grain. For temperatures $T < \Jtot < \TCFM$ each grain is in the FM state with different grains magnetic moments being correlated such that the whole system may experience the FM phase transition.

The exchange coupling constant $J$ has the form~\cite{Landau3,Auerbach}
\begin{equation}
\label{Eq_exch}
J\propto\int\int\Psi^{*}_1(\vec{r}_2)\Psi_2^{*}(\vec{r}_1)\frac{e^2}{\epsilon|\vec{r}_1-\vec{r}_2|}\Psi_1(\vec{r}_1)\Psi_2(\vec{r}_2)d^3r_1d^3r_2.
\end{equation}
Here $\Psi_{1,2}$ is the spatial part of the electron wave function located in the first (second) grain; $\epsilon$ is the average dielectric permittivity. The influence of FE matrix on the exchange integral in Eq.~(\ref{Eq_exch}) is twofold:

i) The $\epsilon$-dependent Coulomb interaction potential.
This interaction, and thus the exchange coupling $J$, decreases with
increasing $\epsilon$ in the vicinity of the paraelectric-ferroelectric transition temperature $\TCFE$.

ii) The $\epsilon$-dependent electron localization length $\xi$, Fig.~\ref{Expl}.
This length depends on the Coulomb energy $E_c$, and thus on the dielectric
permittivity $\epsilon$~\cite{Bel2007review}
\begin{equation}
\label{Eq_LocLen}
\xi=a/\ln(E^{2}_{c}/T^2g_t),
\end{equation}
where $g_t$ is the average tunneling conductance.
It increases with increasing $\epsilon$
leading to larger overlap of the electron wave functions, and thus
to the increase of the exchange coupling $J$.

To summarize, there are two competing mechanisms in Eq.~(\ref{Eq_exch}): with increasing
the dielectric permittivity $\epsilon$ the
intergrain Coulomb interaction decreases, while the
electron wave function overlap increases.

We now estimate the exchange coupling $J$ in Eq.~(\ref{Eq_exch}) using the following form of the
electron wave function
\begin{equation}
\label{WaveFunc}
\Psi_{1,2}(\vec{r}) = C \left\{\begin{array}{l}e^{-\frac{a}{\xi}}, \hspace{0.9cm}|\vec{r}\pm\vec{d}/2|<a, \\e^{-\frac{|\vec{r}\pm\vec{d}/2|}{\xi}},~~|\vec{r}\pm\vec{d}/2|>a. \end{array} \right.
\end{equation}
Here $C = \left(\int|\Psi_{1,2}|^2dV\right)^{-1/2}$ is the normalization constant and $d$ is the
distance between two grain centres. Equation~(\ref{WaveFunc})
describes electrons uniformly smeared inside a grain and decaying exponentially outside
the grain. Substituting Eq.~(\ref{WaveFunc}) into Eq.~(\ref{Eq_exch}) we find the intergrain exchange coupling constant
\begin{equation}
\label{Jcal}
J \sim \frac{1}{\epsilon}\left\{\begin{array}{l} e^{-4d/\xi}, \hspace{1cm} d \gg a \\
e^{-4(d-2a)/\xi}, \hspace{0.3cm} d - 2a \ll a. \end{array} \right.
\end{equation}
In general, the exchange coupling can be estimated as $J\sim (1/\epsilon)e^{-\gamma d/\xi}$, with numerical constant $\gamma\le 4$.
Using Eq.~(\ref{Eq_LocLen}) we find
\begin{equation}
\label{ExchFin}
J=J_0\, \epsilon^{\frac{\gamma d}{a}-1},
\end{equation}
where $J_0>0$ is the exchange coupling for permittivity $\epsilon = 1$. $J_0$ decays exponentially
with increasing the intergrain distance $d$ leading to the decrease of
overall exchange coupling $J$ in Eq.~(\ref{ExchFin}) with increasing the distance $d$. This can be
seen using Eq.~(\ref{Jcal}). The exponent
in Eq.~(\ref{ExchFin}) has a clear physical meaning:
the first term, $\gamma d/a$, is due to $\epsilon$-dependent localization length $\xi$,
the second term ($-1$) is due to $\epsilon$-dependent Coulomb interaction.
These mechanisms compete with each other.

The exchange coupling $J$ in Eq.~(\ref{ExchFin}) depends
on the ratio of grain sizes $a$ and the intergrain distances $d$.
For large intergrain distances, $\gamma d > a$, the exponent of dielectric permittivity
$\epsilon$ in Eq.~(\ref{ExchFin}) is positive leading to the increase of exchange coupling $J$
due to the delocalization of electron wave functions. In the opposite case, of small
intergrain distances, $\gamma d < a$, the exchange coupling $J$ decreases with
increasing of $\epsilon$.
\begin{figure}[t]
\includegraphics[width=0.88\columnwidth]{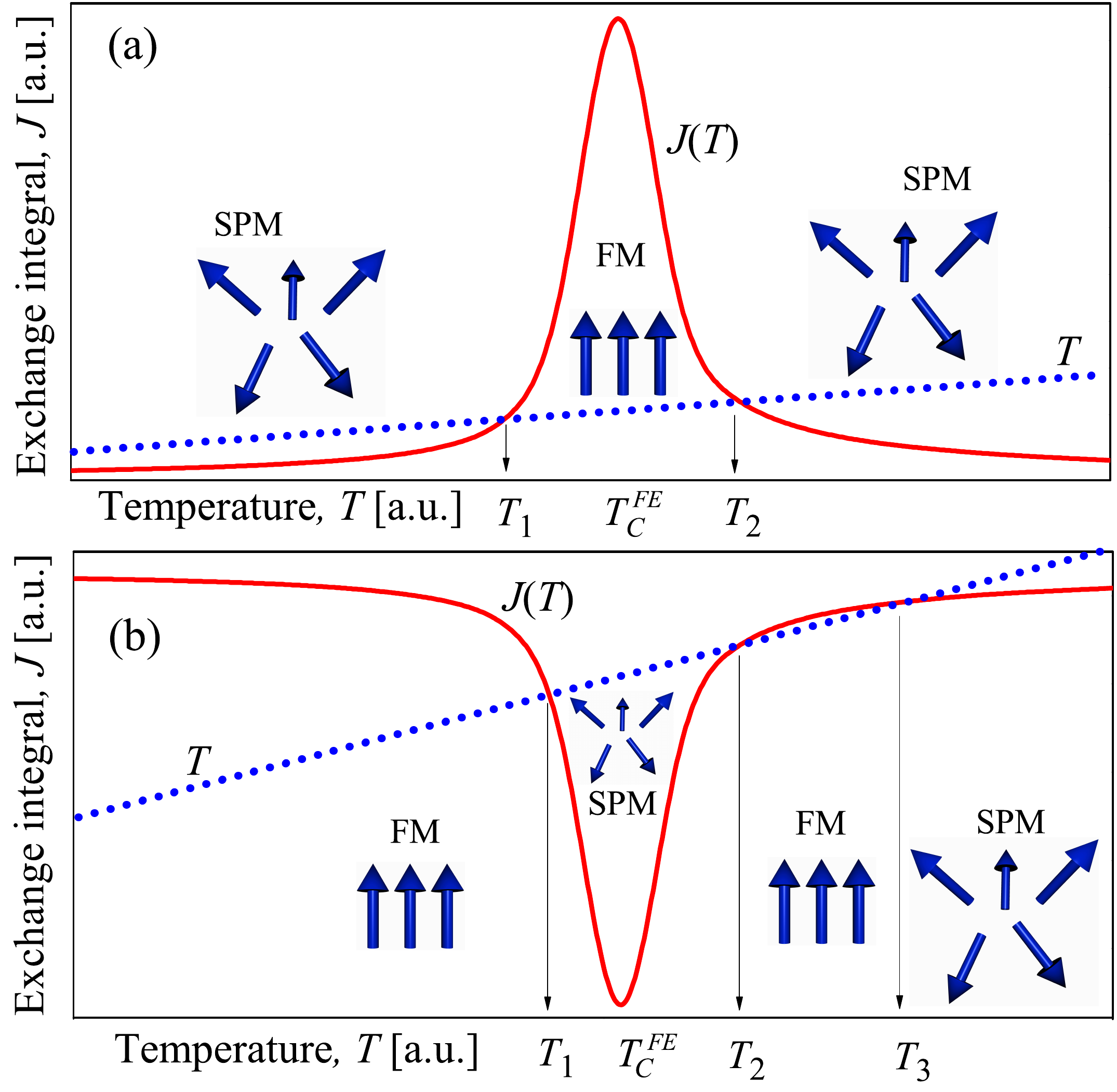}
\caption{\label{Phase1} (Color online) Exchange coupling constant $\JJ$ vs. temperature $T$.
The dotted line stands for temperature with temperatures $T_{1,2,3}$ being the solutions of
Eq.~(\ref{Crit}). (a) Limit of large intergrain distances,
$\gamma d > a$ in Eq.~(\ref{ExchFin}), with superparamagnetic (SPM) state existing
for temperatures $T < T_1$ or $T > T_2$ and the ferromagnetic (FM) state appearing for temperatures
$T_1 < T < T_2$. (b) Limit of small
intergrain distance, $\gamma d < a$ in Eq.~(\ref{ExchFin}), with FM state
appearing for temperatures $T < T_1$ and $T_2 < T < T_3$ and the SPM state being above the
temperature $T_3$ and in the temperature interval $T_1 < T < T_2$.}
\end{figure}

The criterion of SPM - FM phase transition in composite multiferroics can be formulated as follows
\begin{equation}
\label{Crit}
\JJ(\epsilon(\TM))=\TM.
\end{equation}
Here $\JJ$ is the exchange coupling averaged over all pair of grains (it  includes effective nearest neighbor number) and $\TM$ is the transition (or ordering) temperature.

The temperature dependence of the dielectric permittivity $\epsilon (T)$ of composite ferroelectrics --- materials
consisting of metallic grains embedded into FE matrix was discussed recently~\cite{Beloborodov2013, PRB_paper}.
We assume that the metal dielectric constant is very large (infinite) at zero frequency.
Therefore we can write for sample permittivity $\epsilon = \epsilon_{\rm fe} (\Omega/\Omega_{\rm fe})$, where $
\epsilon_{\rm fe}=1+4\pi\chi$ and $\Omega$, $\Omega_{\rm fe}$ being the sample and FE matrix volume, respectively
and $\chi$ is the average susceptibility of FE matrix.

To estimate the dielectric permittivity of FE matrix we consider the region between two particular
neighbouring grains as thin FE film with local polarization perpendicular to the film (grain) boundaries.
The direction of local polarization varies from one pair to another pair of grains, and its
sign is defined by the external and internal electric fields. The origin of
internal field is the electrostatic disorder inevitably present in granular materials.
The behavior of local polarization is described by the Landau-Ginzburg-Devonshire (LGD) theory~\cite{Levan1983,chandra2007landau}.

\section{Discussion.}

Figure~\ref{Phase1} shows the average exchange coupling constant $\JJ$ vs. temperature.
For large intergrain distances, $\gamma d > a$, the exchange coupling
$\JJ$ has a maximum in the vicinity of the ferroelectric Curie temperature $\TCFE$, Fig.~\ref{Phase1}(a).
For small intergrain distances, $\gamma d < a$, the exchange constant $\JJ$
has a minimum, Fig.~\ref{Phase1}(b).
In Fig.~\ref{Phase1} we assume that the grain ferromagnetic Curie temperature is large, $\TCFM > \TCFE$. The
dotted line in Fig.~\ref{Phase1} stands for temperature and the intersections of this line
with exchange coupling curve $\JJ$ correspond to the solution of Eq.~(\ref{Crit}).
The temperatures $T_{1,2,3}$ in Fig.~\ref{Phase1} stand for different ordering temperatures of
SPM - FM phase transitions and correspond to the solution of Eq.~(\ref{Crit}).

\begin{figure}
  \centering
  \includegraphics[width=0.97\columnwidth]{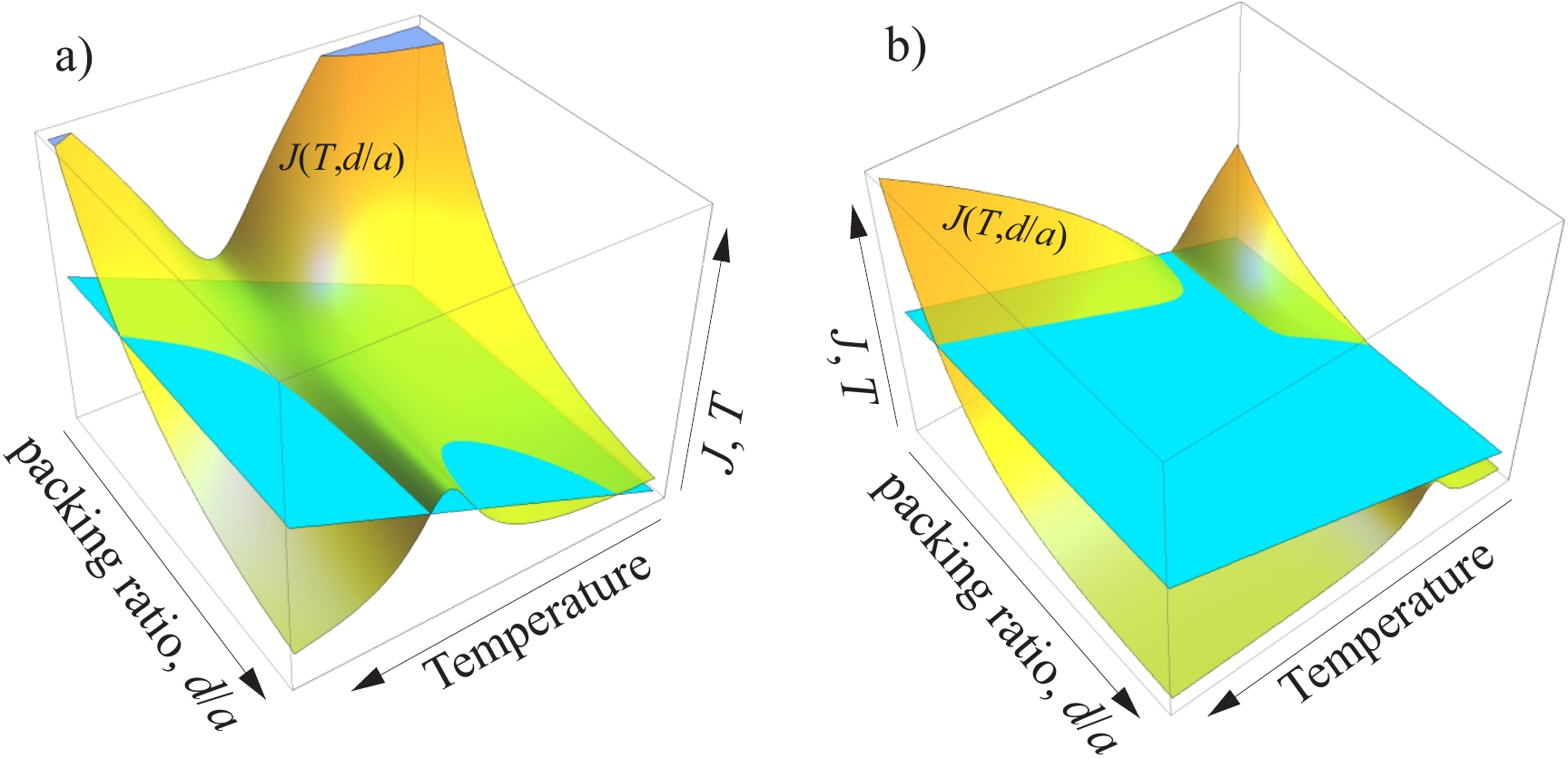}\\
  \caption{(Color online) Graphs a) and b) show the dependence of the exchange coupling $J$
  (complicated surface) in the coordinates packing ratio, $d/a$ and temperature $T$. Plots a) and b) correspond to the limits of
  large and small intergrain distance $d$, respectively. Plane surface stands for temperature $T$. Intersection of
  exchange coupling $J$ and temperature $T$ correspond to the
  magnetic phase transition. The regions with temperature $T > J$ correspond to the SPM
  state, while the region with $T < J$ corresponds to the FM state.
   Following a) and b) we obtain the phase diagrams b) and c)
    in Fig.~\ref{fig1}.}\label{fig4}
\end{figure}
The most interesting region in Fig.~\ref{Phase1} is the intersection of temperature $T$
dotted line with exchange coupling curve, $\JJ$.
For large intergrain distances, $\gamma d > a$ the exchange coupling $\JJ$ exceeds the
thermal fluctuations for temperatures $T_1 < T < T_2$ near the ferroelectric Curie temperature $\TCFE$
leading to the appearance of the global FM state, Fig.~\ref{Phase1}(a).
For temperatures $T < T_1$ or $T > T_2$ the system is in the SPM state.
Interestingly, the FM state appears with increasing the temperature, in contrast to the usual
case where ordering appears with decreasing the temperature. This is related to the fact
that while the magnetic system becomes ordered the FE matrix becomes disordered
with increasing the temperature.

For small intergrain distances, $\gamma d < a$, the exchange coupling $\JJ$
has the opposite behavior, Fig.~\ref{Phase1}(b): The system is in the FM state for temperatures $T < T_1$ and
becomes SPM for temperatures $T_1 < T  < T_2$. Increasing the temperature
the system first experience the transition to the FM state for temperatures $T_2 < T < T_3$
and then goes to the SPM state for temperatures above $T_3$.

Equation~(\ref{Crit}) may not have a solution at any temperatures for small enough
coupling constants $J_0$ in Eq.~(\ref{ExchFin}). In this case
the system will stay in the SPM state.

Figure~\ref{fig4} shows the behavior of intergrain exchange constant $\JJ$
as a function of temperature $T$ and the packing ratio, $d/a$. The flat surface represents
the temperature $T$.
The regions with $\JJ > T$ correspond to the FM state, while the regions with
$\JJ<T$ to the SPM state. Figure~\ref{fig4} was used to obtain the phase
diagrams in Fig.~\ref{fig1}.

To summarize, we obtain the magnetic phase diagram of granular multiferroics
with several phases appearing due to the interplay of ferroelectricity,
magnetism, and strong electron correlations, Fig.~\ref{Phase1}.

\subsection{Requirements for experiment.}

First, we assumed an insulating state of composite multiferroic due to
strong Coulomb blockade,
$E_c \gg \max(T, \JJ)$. The last inequality is not valid in the close
vicinity of the ferroelectric Curie temperature $\TCFE$~\cite{PRB_paper} since
the charging energy $E_c$ is $\epsilon$-dependent and is strongly suppressed in the vicinity of $\TCFE$.
This suppression may lead to the appearance of the metallic state with different criterion of
SPM - FM transition where magnetic coupling between grains
occurring due to electron hopping between the grains~\cite{Tokura1998review,Vons}.
This effect was not considered here.

Above restriction is rather strong and reduces the number of possible FE materials.
The Coulomb gap for 5 nm grains is $E_c = 2000/\epsilon$ K and thus $E_c < 200$ K for dielectric
permittivity $\epsilon > 10$.
In conventional bulk ferroelectrics, such as BTO and PZT, the dielectric permittivity is large, $\epsilon > 100$.
However, in granular materials the thin FE layers are confined by grains leading to
a much smaller dielectric constant~\cite{Zlatkin1998}. Another way to reduce the dielectric constant
is to use the relaxor FE matrix, such as P(VDF-TrFE)~\cite{Tagantsev1998,Zhang2001,Shen2011}.

The origin of magneto-electric coupling in GMF is the long-range Coulomb interaction.
Thus, the magnetic and electric subsystems can be separated in space with FM film
placed above the FE substrate.
This geometry allows using ferroelectrics with large dielectric permittivity. Increasing the
distance between the FE and the FM film one can reduce the influence of FE on
the Coulomb gap.

Second, we assumed that all grains have equal sizes and all intergrain distances are the same.
For broad distribution of grain sizes and intergrain distances the influence of FE matrix on the
exchange coupling constant is smeared. This effect was not taken into account here.

Third, we assumed that the intergrain exchange interaction
is larger than the magneto-dipole interaction and magnetic anisotropy.
This limit is realized in many materials including Ni-SiO$_2$ granular
system~\cite{Boz1972}, where 5 nm Ni grains were embedded into SiO$_2$ matrix with
SPM - FM phase transition observed at temperature
$\TM \approx 300 K \gg \TB$, where $\TB$ is the blocking temperature. Such a high
transition temperature can occur due to the intergrain exchange interaction only.
The results of Ref.~\cite{Boz1972} were repeated for Co-SiO$_2$~\cite{Hel1981}
and Fe-SiO~\cite{Pozd1985} systems with ordering temperature $\TM \approx 300 K$.

Finally, we mention that granular FM show an activation conductivity
behavior supporting the fact that in these materials electrons are localized inside the grains~\cite{Glad1996,Liu2003}. Thus, these
materials can be used for observing the effect discussed in this paper with the proper
substitution of
FE matrix instead of SiO$_2$ matrix.

\subsection{Electric field control of GMF properties.}

The dielectric permittivity of FE matrix can be controlled by the external electric
field in addition to temperature. This opens an opportunity to control the
magnetic state of GMF by the electric field. For example, the magnitude of 
dielectric permittivity of P(VDF/TrFE) ferroelectric relaxor can be doubled by the electric field~\cite{Park2007}. 
According to Eq.~\ref{ExchFin} this leads to four times change 
in the intergrain exchange interaction. This change can cause the magnetic phase 
transition driven by electric field.

\section{Conclusion.}
We studied the phase diagram of composite multiferroics, materials consisting of magnetic
grains embedded into FE matrix,
in the regime of Coulomb blockade. We found that the coupling
of ferroelectric and ferromagnetic degrees of freedom is due to the influence of FE
matrix on the exchange coupling constant via
screening of the intragrain and intergrain Coulomb interaction. We showed that in these
materials the ordered magnetic phase may appear at higher temperatures
than the magnetically disordered phase.

\section{Acknowledgements.}
I.~B. was supported by NSF under Cooperative Agreement Award EEC-1160504 and NSF Award DMR-1158666.

\bibliography{GFM}

\end{document}